\documentclass[aps,twocolumn,showpacs,showkeys,floatfix,longbibliography,superscriptaddress]{revtex4-1}

\usepackage{graphicx}
\usepackage{textcomp}
\usepackage{dcolumn}
\usepackage{amsmath}
\usepackage{amssymb}
\usepackage{epstopdf}
\usepackage{rotating}
\usepackage{color}
\usepackage{natbib}
\usepackage{url}
\setcitestyle{square}
\usepackage{textgreek}
\usepackage{multirow}
\usepackage{makecell}
\usepackage{setspace}
\usepackage{etoolbox}
\usepackage{lipsum}
\usepackage{booktabs}

\begin{document}

\title{Missing data and bias in physics education research: A case for using multiple imputation}

%\pacs{01.40.Fk, 01.40.G--, 01.40.gb I. }

\author{Jayson Nissen}
\affiliation{Department of Science Education, California State University Chico, Chico, CA, 95929, USA}
\author{Robin Donatello}
\affiliation{Department of Mathematics and Statistics, California State University Chico, Chico, CA, 95929, USA}
\author{Ben Van Dusen}
\affiliation{Department of Science Education, California State University Chico, Chico, CA, 95929, USA}

\begin{abstract}
Physics education researchers (PER) commonly use complete-case analysis to address missing data. For complete-case analysis, researchers discard all data from any student who is missing any data. Despite its frequent use, no PER article we reviewed that used complete-case analysis provided evidence that the data met the assumption of missing completely at random (MCAR) necessary to ensure accurate results. Not meeting this assumption raises the possibility that prior studies have reported biased results with inflated gains that may obscure differences across courses. To test this possibility, we compared the accuracy of complete-case analysis and multiple imputation (MI) using simulated data. We simulated the data based on prior studies such that students who earned higher grades participated at higher rates, which made the data missing at random (MAR). PER studies seldom use MI, but MI uses all available data, has less stringent assumptions, and is more accurate and more statistically powerful than complete-case analysis. Results indicated that complete-case analysis introduced more bias than MI and this bias was large enough to obscure differences between student populations or between courses. We recommend that the PER community adopt the use of MI for handling missing data to improve the accuracy in research studies.
\end{abstract}
\maketitle

\section{Introduction}
Physics education research (PER) commonly handles missing data by using complete-case analysis (a.k.a. listwise deletion, casewise deletion, and matched data) \citep{Nissen2018a,Nissen2018b}. Complete-case analysis removes any individuals who are missing any data from the analysis. This method is common because it is easy to implement. However, discarding data lowers the statistical power of the analysis and may bias the results \citep{Schafer1999,Little2014,Cheema2014,Donner1982}. 

\par
\begin{figure}
\includegraphics[width=1\columnwidth]{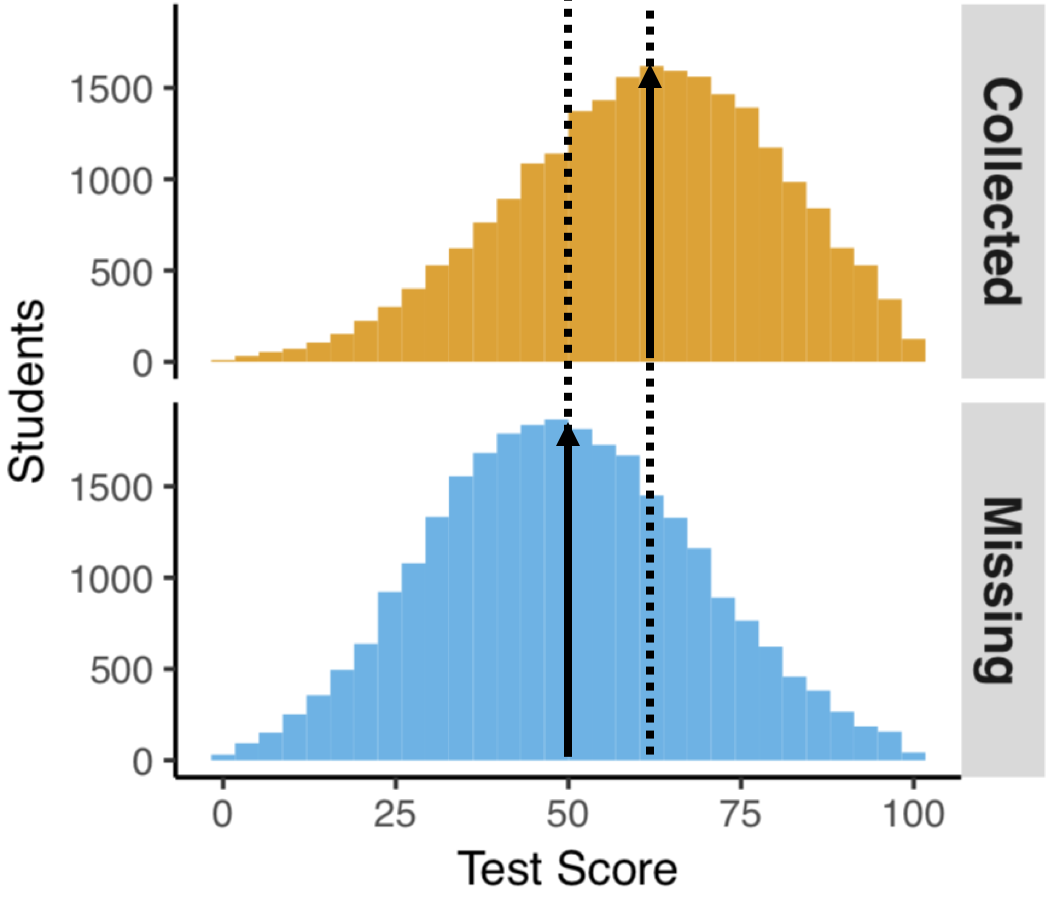}
\caption{Simulated distributions of missing and collected data with means indicated to illustrate data that is not MCAR.}
\label{fig:Not MCAR}
\end{figure}

	Complete-case analysis produces reliable results so long as the missing data is missing completely at random (MCAR) \citep{Schafer1999}. For MCAR, the missingness is completely independent of any observed or missing data \citep{Rubin1976}. We are not aware of any studies in PER that have explicitly tested the MCAR assumption. \citet{VanNess2007} and \citet{Fielding2009} provide examples of these tests in epidemiology and health research. The few studies that have explicitly compared participants and non-participants using course grades \citep{Kost2009,Kost-Smith2010,Nissen2016,Nissen2018b} all indicate that students with higher course grades are more likely to provide complete data. Students with higher course grades also tend to do better on concept inventories and attitudes surveys \citep{Nissen2018b}. PER studies that use these instruments likely do not meet the MCAR assumption because the missing data disproportionately comes from students with lower grades who tend to have lower scores. Therefore, as illustrated by the simulated data in Fig.~\ref{fig:Not MCAR}, the distribution of the collected data and the missing data likely differ. This difference may create biased results. For example, on concept inventories the mean scores will be higher if the data mostly comes from students that earned As and Bs than if it comes from all of the students. 

\par 	As participation rates drop, the skew in representation toward students who receive higher grades typically increases \citep{Nissen2018b}. This increased skew in participation tends to raise the size of the difference between the collected and missing data, leading to a greater likelihood of bias in any subsequent analyses. We are not aware of any studies in PER that have investigated this potential bias, how large this bias may be, nor what impact it could have on understanding student learning in college physics courses.

\par 	Multiple imputation (MI) \citep{Rubin1987} handles missing data without discarding any values by imputing the missing values using statistical models based on the available data. MI completes this process $m$ times to create $m$ complete data sets, analyzing each of those complete data sets with traditional methods to produce $m$ results, and combining the $m$ results into a single mean, variance, and standard error for each of the statistics being calculated. MI \citep{Rubin1996} provides a consistently superior alternative to complete-case analysis. Research shows that MI has greater statistical power and less biased results than complete-case analysis \citep{Schafer1999,Cheema2014,Pampaka2016,Dong2013}. This superior performance results from MI not relying on the assumption that the data is MCAR and from MI using all of the available data to build accurate and reliable models. A search of the Sage journals for the term `multiple imputation' during the preparation of this manuscript indicated that education researchers use of MI as the search identified 2,876 research articles on education that referenced MI. A similar search of the Physical Review database for the term `multiple imputation' identified only four studies in PER that referenced the term. Of these four studies, only two used MI \citep{Dou2016,Nissen2018a}, and we only know of one other PER article outside of Physical Review that used MI \citep{Nissen2018b}. 

\section{Research Question}

In this article, we compare and contrast the bias introduced by using either complete-case analysis or MI to analyze concept inventory data with participation skewed toward higher performing students. We designed the study to cover a broad range of variables we identified as pertinent to concept inventory data. The results inform how likely complete-case analysis biases results in the PER literature and the possible size of those biases. By comparing complete-case analysis and MI we hope to raise awareness in the PER and discipline based education research communities about methods for handling missing data in quantitative studies. 
\par 	To compare the accuracy for complete-case analysis and MI we examined the following research question:

\begin{itemize}
\item When controlling for the relationships between grade, concept inventory scores, grade distributions in a course, and participation rates, to what extent do complete-case analysis and MI produce biased results for posttest scores? 
\end{itemize}

\par 	If the results indicate that complete-case analysis provides inaccurate results compared to MI, these results could motivate researchers to use MI in their studies. The results could also provide reviewers and editors with a resource to push against the use of complete-case analysis and to push for improved reporting and transparency about data collection and analysis in future studies.

\section{Literature Review}

\subsection{Missing data in PER studies}

To inform the common research practices around reporting and handling missing data, we reviewed the published literature in the American Journal of Physics and in Physical Review --  Physics Education Research. We identified 28 studies that reported pretest and posttest scores for concept inventories in introductory physics courses. We did not include studies that used either pretest or posttest scores but did not report descriptive statistics for student performance. Of these 28 studies, six provided adequate descriptive statistics to calculate the participation rates and one \citep{Cahill2018} stated the range of participation rates across the courses sampled in the study, as shown in Table \ref{tab:lit review}. The participation rates ranged from a low of 30\% to a high of 80\%. 

\par 	Twenty-three of the studies we reviewed used complete-case analysis. For studies that did not report how they handled missing data, we inferred from the matched number of pretests and posttests that the researchers used complete-case analysis. Five studies calculated descriptive statistics using all available data. These 28 studies do not include the three studies in PER that used MI, which we discussed earlier. We excluded these three articles from the 28 studies that we reviewed because two of them did not report pretest and posttest scores on concept inventories \citep{Nissen2018a,Dou2016} and we discuss the third article \citep{Nissen2018b} below.
\par 	Only three of the seven studies that reported participation rates, shown in Table \ref{tab:lit review}, provided average grade data for the participants and non-participants. All three studies disaggregated the data by gender. The participants in these three studies had much higher grades than the students who did not participate in the study, with a B- on average for participants and a C on average for nonparticipants. These differences in grades indicate that the missing data in these studies does not meet the assumption of MCAR required for complete-case analysis. The under representation of low-performing students raises the possibility that the results reported in these studies were positively biased.

%Table 1
\begin{table*}[t]
\caption{Participation rates and descriptive statistics for students' grades from prior studies published in Physical Review Physics Education Research. Descriptive statistics include mean ($\mu$), sample size ($N$), and standard deviation ($\sigma$). Grades are in GPA units on a 0 to 4 scale.}\label{tab:lit review}
{\scriptsize
\begin{tabular}{lcccccp{.1cm}ccccc}\hline\hline\noalign{\smallskip}
\multirow{2}{*}{Study}&\multirow{2}{*}{Instruction}&\multirow{2}{*}{~Gender~}&\multicolumn{3}{c}{Participant grades}&&\multicolumn{3}{c}{Nonparticipant grades}&&Participation\\\cline{4-6}\cline{8-10}\noalign{\smallskip}
&&&$\mu$&$N$&$\sigma$&&$\mu$&$N$&$\sigma$&&Rate\\
\hline\noalign{\smallskip}
\multirow{2}{*}{Nissen, 2016 \citep{Nissen2016}}&\multirow{2}{*}{Active}&Male&2.69&90&1.28&&2.1&92&1.28&&0.49\\
&&Female&2.78&27&1.26&&2.05&13&1.16&&0.68\\
\multirow{2}{*}{Kost-Smith, 2010 \citep{Kost-Smith2010}}&\multirow{2}{*}{Active}&Male&2.85&1257&0.8&&1.93&500&1.1&&0.72\\
&&Female&2.80&447&0.8&&1.96&114&1.2&&0.80\\
\multirow{2}{*}{Kost, 2009 \citep{Kost2009}}&\multirow{2}{*}{Active}&Male&2.82&1563&0.8&&2.14&1152&1.2&&0.58\\
&&Female&2.74&533&0.8&&1.89&315&1.1&&0.63\\
\multirow{2}{*}{Henderson, 2017 \citep{Henderson2017}}&\multirow{2}{*}{Lecture}&Male&-&1084&-&&-&342&-&&0.76\\
&&Female&-&323&-&&-&102&-&&0.76\\
\multirow{2}{*}{Brewe, 2010 \citep{Brewe2010}}&Modeling&All&-&258&-&&-&64&-&&0.8\\
&Lecture&All&-&758&-&&-&1743&-&&0.3\\
\multirow{2}{*}{Cahill, 2014 \citep{Cahill2014}}&Lecture&All&-&366&-&&-&314&-&&0.54\\
&Active&All&-&773&-&&-&448&-&&0.63\\
\multirow{2}{*}{Cahill, 2014 \citep{Cahill2014}}&Lecture&All&-&360&-&&-&219&-&&0.62\\
&Active&All&-&738&-&&-&384&-&&0.66\\
Cahill, 2018 \citep{Cahill2018} &Both&All&-&-&-&&-&-&-&&0.34 -0.59\\
\noalign{\smallskip} \hline\hline
\end{tabular}
}
\end{table*}

%Figure 2
\begin{figure*}
\includegraphics[width=.8\linewidth]{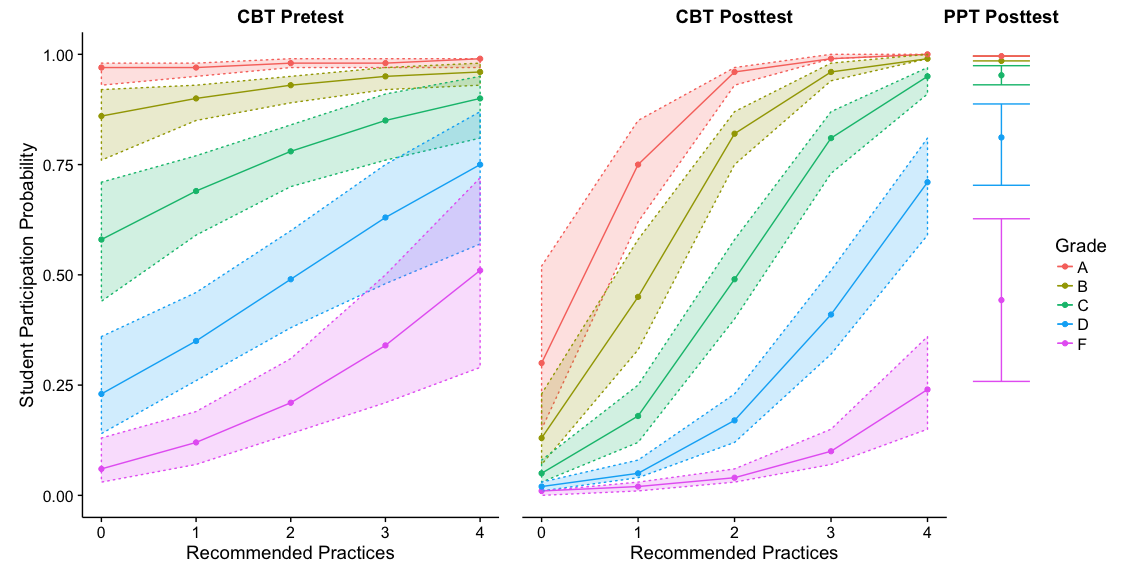}
\caption{Participation rates for computer-based tests (CBT) and paper- and pencil tests (PPT) from \citet{Nissen2018b}. Participation on the PPT pretest is not shown because it closely clustered around 100\% for all grades. Recommended practices measured the total number of up to four actions instructors could take to motivate students to participate in the CBTs: participation credit on the pretest, participation credit on the posttest, in class reminders, and email reminders.}
\label{fig:lonestar}
\end{figure*}

\subsection{An Investigation of Participation on Low-Stakes Assessments}
\citet{Nissen2018b} used an experimental design to investigate the differences in performance and participation on paper-and-pencil tests (PPT) administered in class and computer-based tests (CBT) administered online outside of class. In this article, we focus on their participation models. Data for the study came from 1,310 students in 25 sections of 3 different introductory physics courses at one institution. Instructors asked every student to complete four assessments: paper- and computer-based pretests and posttests. Instructors reported using four different practices to motivate students to participate: participation credit on the pretest, participation credit on the posttest, in class reminders, and email reminders. They modeled the participation rates of the students using Hierarchical Generalized Linear Models to produce estimates of the likelihood that students would provide data on the low-stakes assessments. The hierarchical models nested the data in three levels: tests nested in students who nested in course sections. Variables in the final model included paper pretest, computer pretest, paper posttest, and computer posttest at the test level; final course grade and gender at the student level, and participation practices treated as a continuous variable from 0-4 based on the total number of practices instructors used at the course section level. The coefficients in generalized linear models are the log of the odds ratio, e.g. logits. Because logits are uncommon, nonintuitive, and beyond the scope of this article, we will focus on the predicted participation rates reported by Nissen and colleagues, which are shown in Fig. \ref{fig:lonestar}. 

\par Nissen and colleagues found participation tended to be higher on pretests than on posttests, participation tended to be higher on paper-and-pencil tests than on computer-based tests, and students that earned higher grades participated at higher rates than those that earned lower grades. The final model predicted that participation on computer-based tests matched that on paper-and-pencil tests when instructors used all four practices to motivate student participation. The differences in participation across student grades existed no matter what practices instructors used to motivate their students to participate. 
\par    Their final model predicted female students participated at slightly higher rates than male students, but this difference was not statistically significant. To generate the participation rates represented in Fig. \ref{fig:lonestar}, Nissen and colleagues input the mean value for gender into their participation model. 

\subsection{Summary of Missing Data in PER Studies}
\par Higher participation rates for higher achieving students occurred in all of the studies that we reviewed that reported information on participation. We cannot rule out the possibility that only studies with a skew in participation reported on differences in grades between participants and non-participants. However, \citet{Kost-Smith2010} reported one of the highest participation rates and reported this skew while \citet{Nissen2018b} found that the skew became smaller as the participation rate increased. Furthermore, \citet{Nissen2018b} tested for the relationship between grade and participation because it was reported in earlier studies \citep{Nissen2016,Kost-Smith2010,Kost2009}. Until studies show no relationship between course grades and participation, the literature consistently and reliably indicates that students who earn higher grades are more likely to participate than those that do not.
\par The positive relationship between grade and participation indicates that concept inventory data is not MCAR. This consistent failure to meet the assumptions necessary for complete-case analysis to produce accurate results combined with the almost exclusive use of complete-case analysis raises the possibility that results in PER studies that use pre-post concept inventories are positively biased to varying extents. 

%Figure 3
\begin{figure*}
\includegraphics[width=1\linewidth]{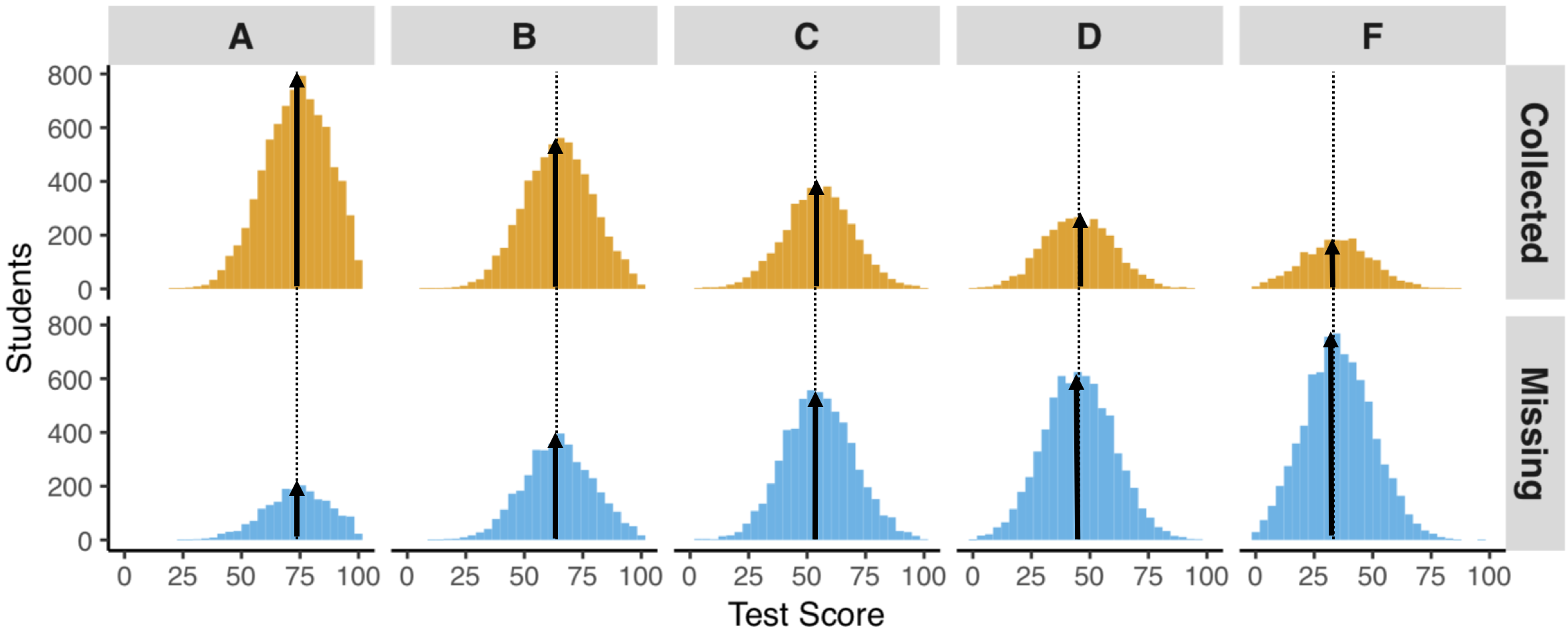}
\caption{The simulated concept inventory data shown in Fig. 1 disaggregated by student's course grades. The similar distribution for each grade indicates that the data is MAR because course grade accounts for the missingness. We made the course grades follow a flat distribution ($N_A=N_B=N_C=N_D=N_F $) to focus the differences between the collected and missing data on the similar distribution by grade that indicates the data is MAR and to illustrate how combining the data results in Fig. 1, where the collected and missing distributions differ.}
\label{fig:MAR}
\end{figure*}

	\subsection{Types of missing data}
The statistical methods underlying complete-case analysis assumes the data is MCAR. MI makes no explicit assumption about the missingness of the data, however many software packages implementation of MI assumes missing at random (MAR) data. \citet{Rubin1976} coined three terms to classify the relationships between the mechanisms of the missingness and the missing and observed values themselves. 

\begin{itemize}
	\item Missing completely at random (MCAR): all of the cases have the same probability of being missing. No relationship exists between the probability of a case being missing and any values in the dataset. This assumption can be partially tested \citep{Little1988}.
	\item Missing at random (MAR): The missingness is independent of the value of the missing data but is conditionally dependent on other observed variables that can explain all of the missingness. For example, a researcher has blood pressure, age, and cardiovascular disease data. They are concerned that the blood pressure data is not missing at random because older people with cardiovascular disease are more likely to report their blood pressure than young healthy people. Provided the age and cardiovascular disease data can explain the missingness in the data, the data is MAR. 
	\item Missing not at random (MNAR): The missingness depends on both the observed and unobserved data. For example, wealthy and poor people who chose not to report their income for fear of being stigmatized due to their income. Since the reported variable is related to the likelihood of reporting and no other variable can explain the missingness, the data is MNAR.
\end{itemize}

\par 	In real world data, the boundary between MAR and MNAR cannot be firmly established because doing so requires observing the unobserved data. Instead, researchers must make reasonable arguments to evaluate the mechanism of missingness. Simulation studies like the one we present in this manuscript allow researchers to build models with data that is known to be missing based on one of the three missingness classifications.
\par 	\citet{Bhaskaran2014} provide a brief article explaining MAR. They argue \citep[p.~1337]{Bhaskaran2014}, ``... the terminology describing missingness mechanisms is undeniably confusing. In particular, `missing at random' is often conflated with `missing completely at random', leading researchers to mistakenly conclude that any systematic patterns or mechanisms underlying the missing data contraindicate the use of multiple imputation.'' We adapted the following scenario from Bhaskaran and Smeeth's article to present MAR in a common context for PER. Their article provides a more thorough discussion of MAR.
\par 	We present the following scenario as an example of MAR. A research team collected concept inventory data, but they are concerned that the data is MNAR because the students who participated had much higher grades than the students who did not participate. 
Fig.~\ref{fig:Not MCAR} illustrates this scenario.  The researcher can use the grade data to argue that the data is MAR because the missingness in the concept inventory data can largely be explained by the students grades, as illustrated by Fig.~\ref{fig:MAR}. In the case of MAR data, splitting the data in Fig.~\ref{fig:Not MCAR} by grade results in Fig.~\ref{fig:MAR} and shows similar distributions between collected and missing data for each grade. The distribution of missing data for the A students looks similar to the complete data for the A students and so on for each group of students. The researcher can argue that within each group of students (A, B, C, D, and F) the primary factors related to their participation were not related to their performance (i.e., traffic, illness, a death in the family, etc.) and the groups with lower participation had more of these unrelated events overall. The difference in the aggregated data, Fig.~\ref{fig:Not MCAR} resulted from the difference in the proportion of students that participated for each grade, which is illustrated by the height of the histograms in Fig.~\ref{fig:MAR}. 

\subsection{The persistence of complete-case analysis}
Despite the known and proven bias caused by ignoring missing data when it is not MCAR, many research fields continue to use complete-case analysis. \citet{Cheema2014} points out that complete-case analysis and other error prone methods for handling missing data are common in education research. \citet{King2001} found that 94\% of political scientists used complete-case analysis, resulting in losing one third of their data on average. In biomedical research, few studies accurately report the amount of missing data or how they handled it, and those that do most commonly report using complete-case analysis \citep{Fernandes2011,Burton2004,Horton2007,Masconi2015}. These four critiques of complete cases analysis in biomedical research span from 2004 to 2015, indicating that researchers can consistently critique the use of complete-case analysis with little improvement in a field's practices. 

\subsection{Imputation of missing data}
Imputation is a principled technique for handling missing data \citep{Little2014}. Imputation fills in the missing data with plausible values, such that a researcher can analyze the now complete data set without concern for missing data. Imputation methods fall into two broad categories: deterministic and probabilistic. We focus on probabilistic imputation methods in this article, but provide a brief review of deterministic methods for contrast.

\par \emph{Deterministic options} for imputation include mean imputation and last observation carried forward. Mean imputation replaces the missing values with the mean value for that variable. Researchers use last observation carried forward with longitudinal data to replace the missing data with the last observed value for all subsequent measurements. Both are problematic because they (1) do not preserve the relationships between variables and (2) as with any single imputation approach, do not account for the error incurred by the imputation process itself. These deterministic methods treat the missing values as if they were known, which can lead to inappropriately small variances and an erroneously increased chance of statistically significant findings \cite{Malhotra1987}.

\par
\begin{figure}
\includegraphics[width=1\columnwidth]{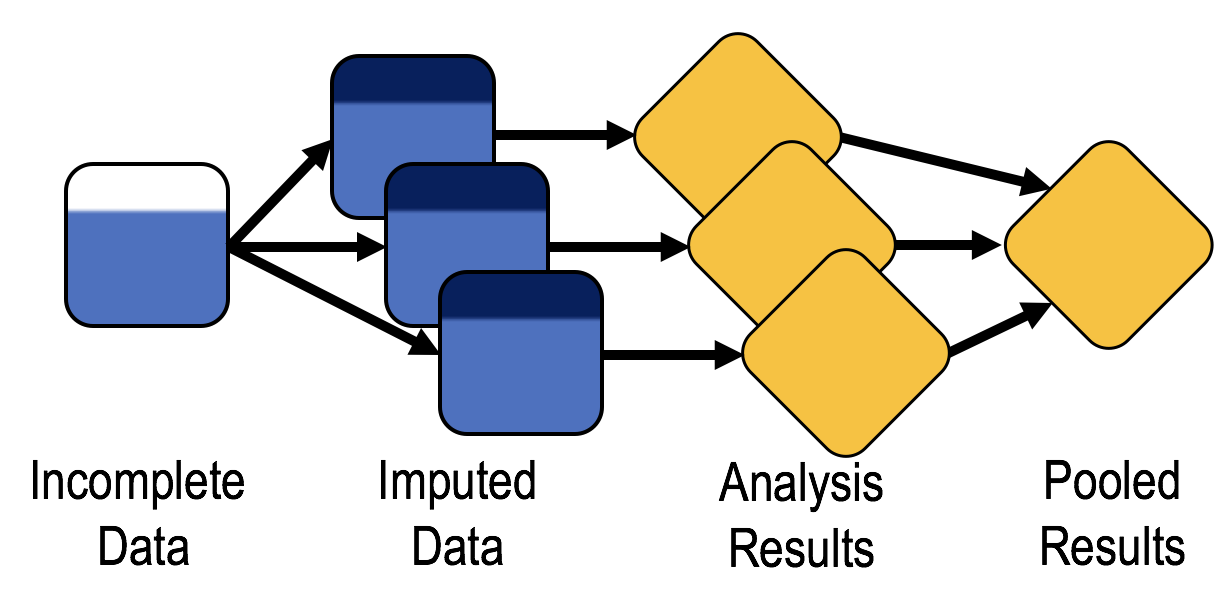}
\caption{The multiple imputation (MI) process. In the first step missing data (shown in white) is imputed (shown in dark blue) to create $M$ complete data sets, with $M=3$ shown here. Then each complete imputed dataset is analyzed using standard methods such as linear regression. Finally the results are pooled using Rubin's Rules.}
\label{fig:MI}
\end{figure}

 	\emph{Probabilistic options} for imputation include multiple imputation (MI) and maximum likelihood estimation. In this article, we demonstrate the use of MI \citep{Little2014} because it is a probabilistic approach for addressing missing data across a wide range of applications \citep{Schafer1999} and because research finds that MI is more statistically powerful and more accurate than other methods for handling missing data \citep{Cheema2014,Dong2013}. The idea behind MI is graphically presented in Fig.~\ref{fig:MI}. The first step applies an imputation procedure containing a random component (such as predictive mean matching, which is described below) to a dataset with missing data $M$ times to generate different imputed values for each piece of missing data and generate $M$ complete data sets. Step two calculates the desired estimate from the analysis, such as a mean or regression coefficient, on each data set separately using standard analytical methods. The final step pools the estimates using simple combining rules, also known as \emph{Rubin's Rules} \citep{Rubin1987}, which are described later in Eqs.~(1-5). These pooled results then properly reflect the variation in the original estimates and the variation introduced by the imputation process itself. 

%Figure 4

\par 	The plausibility of the imputed values generated in the first step relies entirely on the model used for the imputation. Simplistic imputation models that do not use information contained in related variables will impute values that are not an accurate reflection of what the missing data could have been. For example, imputation models need to account for whether the data is longitudinal or if there is reason to suspect the data is MNAR, and the models need to include known correlations and relationships between variables or measures. In short, MI is only as good as the imputation model being used to create the imputed values. 
\par 	Many software programs have built in or add-on methods to perform MI, both the imputation and pooling steps.  In this paper we used the MICE \citep{vanBuuren2011} package in RStudio V.~1.1.456 \cite{R2018}. The MICE package uses predictive mean matching, an imputation method developed by \citet{Little1988} in 1988, as the default model to impute missing data for continuous variables. Predictive mean matching uses the following process \citep{Vink2014} to multiply impute the missing data based on the data the researcher collected. We use -hat (~$\hat{}$~) to differentiate observed $y$ and predicted $\hat{y}$ values.

\begin{enumerate}
\item Using the portion of the data with no missing values, build a linear model ($b$) by calculating the least squares estimates of the regression coefficients $\hat{\beta}$, the model residuals $\hat{\epsilon}$, and variance of the residuals $\hat{\sigma}$. 
\item Create a new linear model ($b^{(m)}$) by randomly drawing values for the regression coefficient from a probability distribution centered on $\hat{\beta}$ with variance derived from $\hat{\sigma}$ and $\hat{\epsilon}$. 

\item Use \emph{b} to generate predictions $\hat{y}_{i}$ for all cases with fully observed data, and $b^{(m)}$ to generate predicted values $\hat{y}_{j}^{*}$ for all cases with missing data ($ i \neq j$). 

\item For each case with a missing value, identify a set of $k$ predictions on observed data ($\hat{y}_{i}$) that are close to the predicted value $\hat{y}_{ j  }^{*}$. The $k$ observed values $y_{i}$ from these matched records form a donor pool of values, where k should vary between 3 and 10 depending on the size of the complete data set. The MICE package uses $k=5$.

\item Randomly choose one observed value $y_{i}$ from the donor pool to impute the missing value.  

\item Repeat steps 2-5 for each of the $M$ imputations.

\end{enumerate}

\par 	Following analysis of each complete dataset researchers, with the aid of statistical software, pool the individual results from across the $M$ imputations using Rubin's Rules to generate valid estimates and intervals of the quantities of interest. To explain Rubin's Rules, let $\delta$ be the parameter whose estimate we desire to obtain from an analysis (i.e., a mean, correlation, or regression slope). Given $M$ imputed data sets, $M$ estimates of $\delta: (\hat{\delta}_{1},\hat{\delta}_{2},\ldots, \hat{\delta}_{M})$ are generated and used to calculate the following quantities. 
\begin{itemize}
\item The overall estimate of the parameter is the average of the individual point estimates. 
\begin{align} \label{eq:rubinsrules}
	\hat{Q} & = \frac{1}{M}\sum_{m=1}^{M} \hat{\delta}_{m}. \\
\intertext{\item The within-imputation variance is the average of the individual variances.}
	U & = \frac{1}{M}\sum_{m=1}^{M} Var(\hat{\delta}_{m}). \\
\intertext{\item The between-imputation variance is the variance of the estimates} 
	B & = Var(\hat{\delta}_{1},\hat{\delta}_{2},\ldots, \hat{\delta}_{M}). \\
\intertext{\item The total variance is a weighted average of the within and between imputation variances.}
	T &= U + (1+\frac{1}{M})B,\\
\intertext{\item And, 95\% intervals are calculated using the total variance.}
	\hat{Q} &\pm 1.96*\sqrt{T}.
\end{align}
\end{itemize}

\par 	The resulting variance of the combined estimate then accounts for both the within and between data set variances. The predictive mean matching process incorporates randomness in steps 2 and 5. The amount of variance introduced in these steps depends on the variability and size of the data set being modeled. If the linear regression in step 1 provides an excellent fit with small standard errors for the coefficients, then little variability will be added by step 2 because each of the $M$ linear models will be very similar and thus will generate similar predictions across the $M$ imputations. Step 5 adds little variability if the data set is large because a large data set will likely have several similar values that will populate the donor pool. By pooling the within and between imputation variances, Rubin's Rules provides standard errors for the estimates based on all of the available information that account for the uncertainty introduced by the missing data.

\subsection{Comparisons of methods for handling missing data in education research}

\citet{Pampaka2016} compared complete-case analysis to MI for handling missing data using a dataset that originally had large portions of missing data that they were able to fill in with subsequent data collection. This design allowed them to compare the results for MI and complete-case analysis of the missing data to the true values for the dataset with no missing data. The total dataset included 1,374 students, but complete-case analysis reduced the data to 495 students. Pampaka and colleagues used a logistic regression to model the probability that students dropped out of the current mathematics course they were enrolled in. The model included predictor variables for the mathematics course students took before this course, student's disposition towards math, student's math self-efficacy, and student's grade on the General Certificate of Secondary Education (GCSE) for mathematics. Students who received an A on the GCSE were three times more likely to provide data than students who received a C, indicating that the data was not MCAR. Both the complete case and MI models provided similar relationships between the variables to those in the true models. However, MI produced smaller standard errors than complete-case analysis. They concluded that MI provided a much closer approximation of the true values than complete-case analysis. Pampaka and colleagues do not discuss why the complete-case analysis and MI provided similar results or the implications of those similarities, nor does their study provide sufficient details for us to make meaningful inferences about the lack of differences.
\par	\citet{Cheema2014} used a simulation study and two real datasets to provide guidance for researchers in designing studies to account for sample size, proportion of missing data, method of analysis, and method for handling missing data. The analysis compared four methods for handling missing data: multiple imputation, complete-case analysis, mean imputation and maximum likelihood estimation. To characterize the quality of the four methods, Cheema used the root mean square error (RMSE). RMSE is the standard deviation of the results from the multiple simulations about the mean of the results, and is a measure of the random error introduced by the four methods. As such, RMSE does not account for any bias (i.e., systematic error) between the mean of the simulations with missing data and the true values where no data is missing. Cheema compared the four analytical methods across three sample sizes and two levels of missingness. The two levels of missing data were 1\% to 10\% and 11\% to 20\%; very few studies in the PER literature report such low levels of missing data. This design created a decision tree with 24 possibilities. Multiple imputation was the most effective method in 15 cases and maximum likelihood estimation in 7 cases. Similar to \citet{Pampaka2016}, Cheema found that imputation methods increased the statistical power of the studies with samples less than 200 by large enough amounts to warrant the use of imputation methods. Cheema warned that missing data can bias data sets and inferences drawn from studies using these biased dataset. In these cases, he urged researchers to use statistical methods that accounted for that bias. However, Cheema did not measure bias introduced by missing data in his study.

%Figure 4b
\begin{figure*}
\includegraphics[width=1\linewidth]{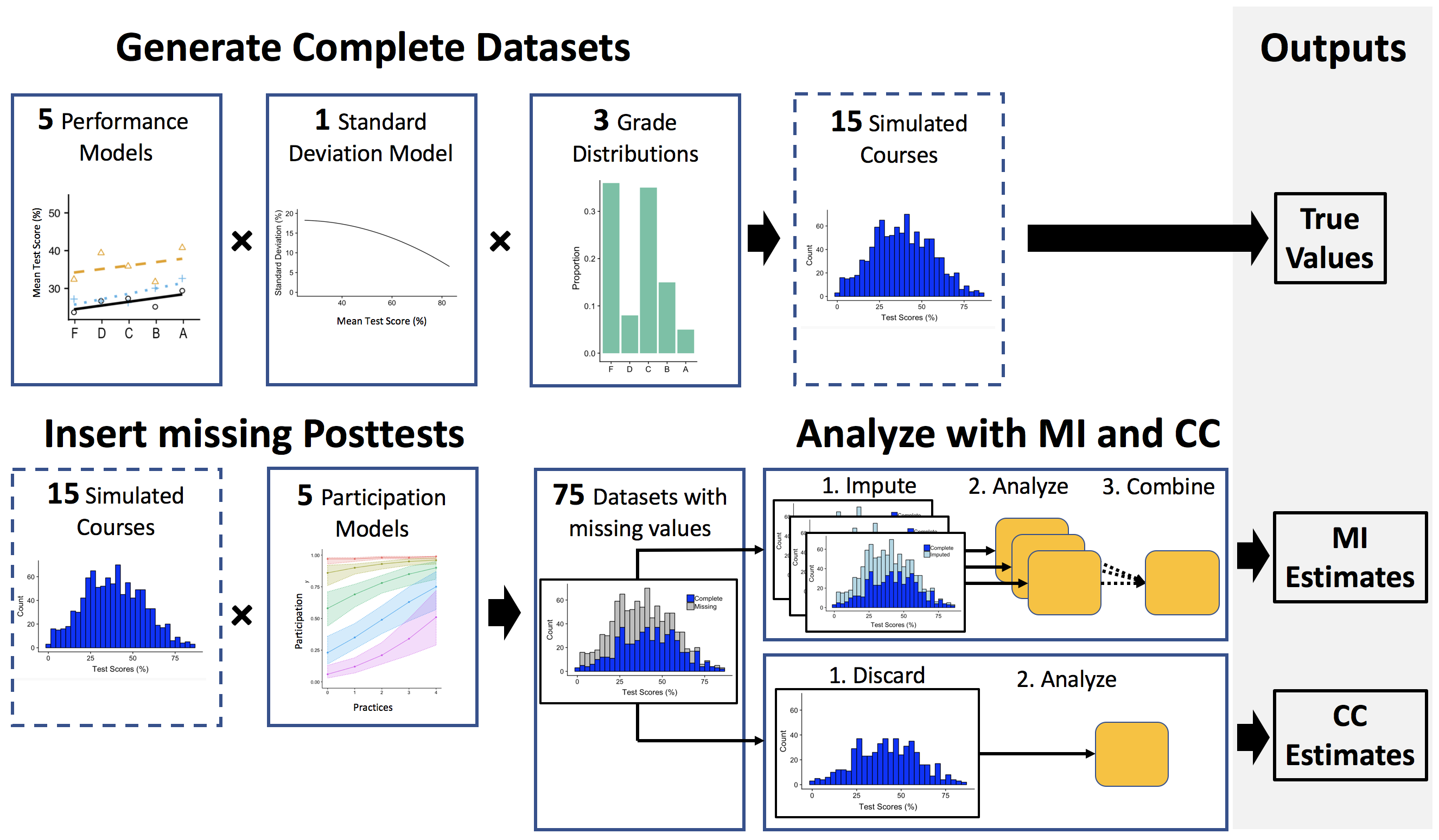}
\caption{Overview of data simulation and analysis methods. In the first stage, we used models of performance, standard deviation, and grade distributions to simulate courses. These simulated courses provided the true values for our analyses. In the second stage, we used participation models to create missing data by deleting posttest scores from the simulated course data. In the third stage, we analyzed the datasets with missing data using both multiple imputation (MI) and complete-case analysis (CC). This stage provided the MI and CC estimates. We used the three outputs (true values, MI estimates, and CC estimates) to investigate the bias introduced by MI and complete-case analysis.}
\label{fig:methods}
\end{figure*}

\par	These two studies illustrate how MI tends to have greater statistical power than complete-case analysis. The trend toward greater statistical power for MI follows from MI using all of the available data and not discarding any data. These studies did not identify bias in the results from either complete-case analysis or MI.

\section{Methods}
We compared the accuracy of estimates from MI and complete-case analysis using simulated course data for grades, pretest and posttest concept inventory scores, and missing values for posttest concept inventory scores. Our analysis focused on course level mean posttest scores as the estimate of interest ($\mu_{post}$). While we focused on posttest means, we also analyzed mean pretest scores ($\mu_{pre}$) because many effect sizes and analytical methods use both pretest and posttest scores. Data simulation included a random component that allowed us to generate complete data, create missing values, and calculate $\mu$ many (20) times to generate a distribution of $\mu$'s. Running the analyses twenty times informed how consistently the measures and methods for handling missing data performed.

\par Figure \ref{fig:methods} illustrates our process for generating the complete and missing data. In the first stage, we simulated complete courses by using five performance models of the relationships between course grades and mean concept inventory scores; one model of the relationship between the mean concept inventory score for a group and the standard deviation of the scores for that group; and three models of grade distributions. This first stage produced the true values ($\mu$) for our analysis. In the second stage, we introduced missing posttest data into the simulated courses using five models of the relationship between participation and course grade based on prior research \citep{Nissen2018b}. Because we removed posttest scores based on course grade, the data was MAR. In the third stage we calculated estimates ($\hat{\mu}$) using complete-case analysis and MI. This design allowed us to assess the effect of the simulation model parameters and the method of handling missing data on the accuracy of the estimates.
\par Because earlier studies did not find large differences in participation rates between male and female students we did not include gender as a variable in our simulated data.

\subsection{Simulating the complete data to generate true results}
\par 	We simulated the course data by simulating data for each of the five course grade subsets (A, B, C, D and F) and then combining the five subsets into a single dataset.  To generate the concept inventory scores, we used a truncated normal distribution, which limited the scores to between 0\% and 100\%. The normal distribution required inputs for mean ($\mu$), standard deviation ($\sigma$), and sample size ($N$). The mean for each grade came from five performance models based on three physics courses investigated by \citet{Nissen2018b}. The standard deviation came from a model of the relationship between the mean and standard deviation for 197 pretest or posttest administrations of concept inventories. The sample size for each grade subset came from the total course size and three grade distributions we developed based on the grade distributions from 192 STEM courses. We used the five performance models and three grade distributions to cover a range of relationships that could occur in PER studies.

\subsubsection{Determining means using the relationships between concept inventory scores and course grades}
\begin{figure}
\includegraphics[width=1\columnwidth]{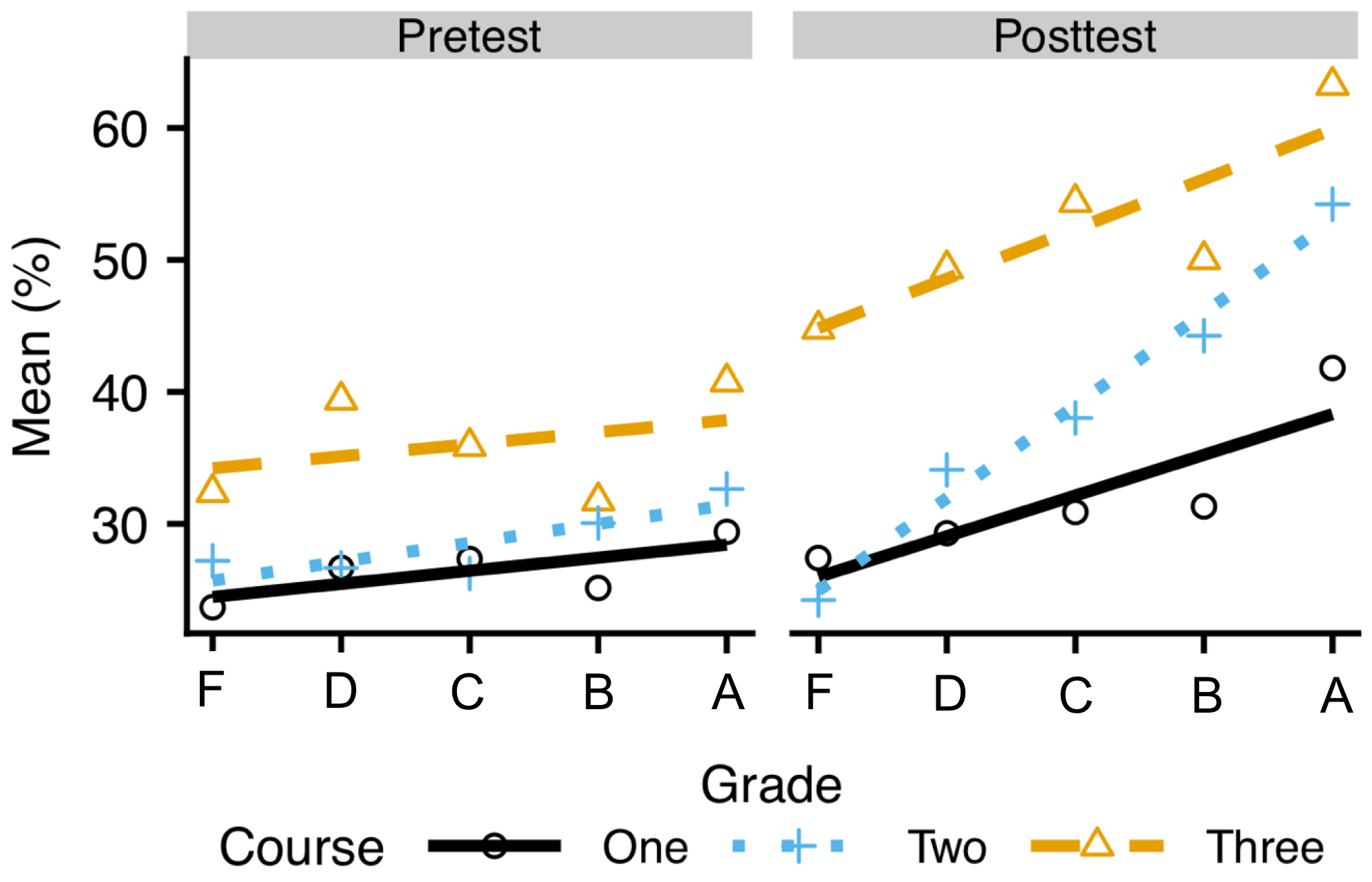}
\caption{Raw data and linear regression fit lines for average pretest and posttest scores for each grade for the three courses described by \citet{Nissen2018b}.}
\label{fig:gradevsciscore}
\end{figure}

To generate realistic concept inventory scores, we examined the relationship between course grade and concept inventory scores using data from \citet{Nissen2018b}. We disaggregated the students in each course by their course grade and calculated the mean concept inventory score for each group of students in each course. We transformed the grades to the numeric values, A=4, B=3, C=2, D=1, and F=0, that the institution used to calculate student grade point average (GPA).  Figure \ref{fig:gradevsciscore} presents the means for each course grade and linear regression fit lines for the pretests and posttests for the three courses. Table \ref{tab:linear models} includes the intercept, slope, and $r^2$ for each linear regression. Based on the scatter plots in Fig.~\ref{fig:gradevsciscore} and the $r^2$ value exceeding 0.5 for 5 of the 6 models, we concluded that a linear model adequately described the relationship between mean concept inventory scores and course grades.

%Figure 7

%Table 2
\begin{table}[t]
\caption{Linear models of the relationship between concept inventory score and course grade for pretest and posttests.}\label{tab:linear models}
{\scriptsize
\begin{tabular}{ccccc}\hline\hline\noalign{\smallskip}
Test&Course&Intercept&Slope&$r^2$\\
\hline\noalign{\smallskip}
Pre		&One	&24.5	&0.99	&0.52\\
Pre		&Two	&25.7	&1.43	&0.69\\
Pre		&Three	&34.2	&0.91	&0.13\\\noalign{\smallskip}
Post	&One	&26.0	&3.08	&0.75\\
Post	&Two	&24.9	&7.02	&0.98\\
Post	&Three	&44.8	&3.77	&0.73\\
\noalign{\smallskip} \hline\hline
\end{tabular}
}
\end{table}

\par 	The mean concept inventory scores represented the average value for each grade about which the models simulated the individual scores. To cover a broad range of performance levels, we built models for five different performance levels that were informed by the linear models from the three courses studied by \citet{Nissen2018b}. The models differ from the results in Table \ref{tab:linear models} because our goal was to cover a broad range of possible relationships rather than to replicate the relationships that we found.  Table \ref{tab:modelparameters} contains the model parameters for the one pretest model and the five posttest models.  Equation (6) shows the generalized equation that we used to calculate the mean score for each grade based on the models in Table \ref{tab:modelparameters}. We started with an \emph{average} model and modified it to create two high-performance models and two low-performance models by varying either the slope or the intercept in the model. The intercept established the mean concept inventory score for the subgroup that earned an F. The slope established the size of the difference between each grade. These five models covered a range of relationships to inform how varying the slope and intercept related to the bias introduced by using MI or complete-case analysis and to provide more robust and generalizable results.

\begin{equation}
\label{eq:9}
\mu_{Grade} = Intercept + Slope*Grade.
\end{equation}

%Table3*
\begin{table}[t]
\caption{Model parameters used to simulate pretest and posttest score data.}\label{tab:modelparameters}
{\scriptsize
\begin{tabular}{lcc }\hline\hline\noalign{\smallskip}
Model&Intercept&Slope\\\hline\noalign{\smallskip}
Pretest&25&2\\
Average&43&6\\
Low Int. &25&6\\
High Int. &58&6\\
Low Slope &43&3\\
High Slope &43&10\\
\noalign{\smallskip} \hline\hline
\end{tabular}
}
\end{table}

\subsubsection{Determining standard deviation using distribution of concept inventory scores}
\par	We used 197 means and standard deviations from either pretests or posttests to build a quadratic model for the relationship between mean and standard deviation. This data came from both the literature and concept inventories collected with the LASSO platform \citep{LASSO}. A quadratic model fit the data because the standard deviation should approach 0 at both of the boundaries of the test scores (0\% and 100\%). Equation 7 describes the fit line. We determined that the quadratic fit line was adequate because the adjusted $r^2$ for the fit line was 0.34, all coefficients were statistically significant with $p<0.001$, and visualizations indicated that the quadratic fit line was appropriate. 

\begin{equation}
\label{eq:8}
\sigma = 16.6+14.6*\mu - 32.2*\mu^2.
\end{equation}

\subsubsection{Determining sample size based on grade distributions in STEM courses}
\begin{figure}
\includegraphics[width=1\columnwidth]{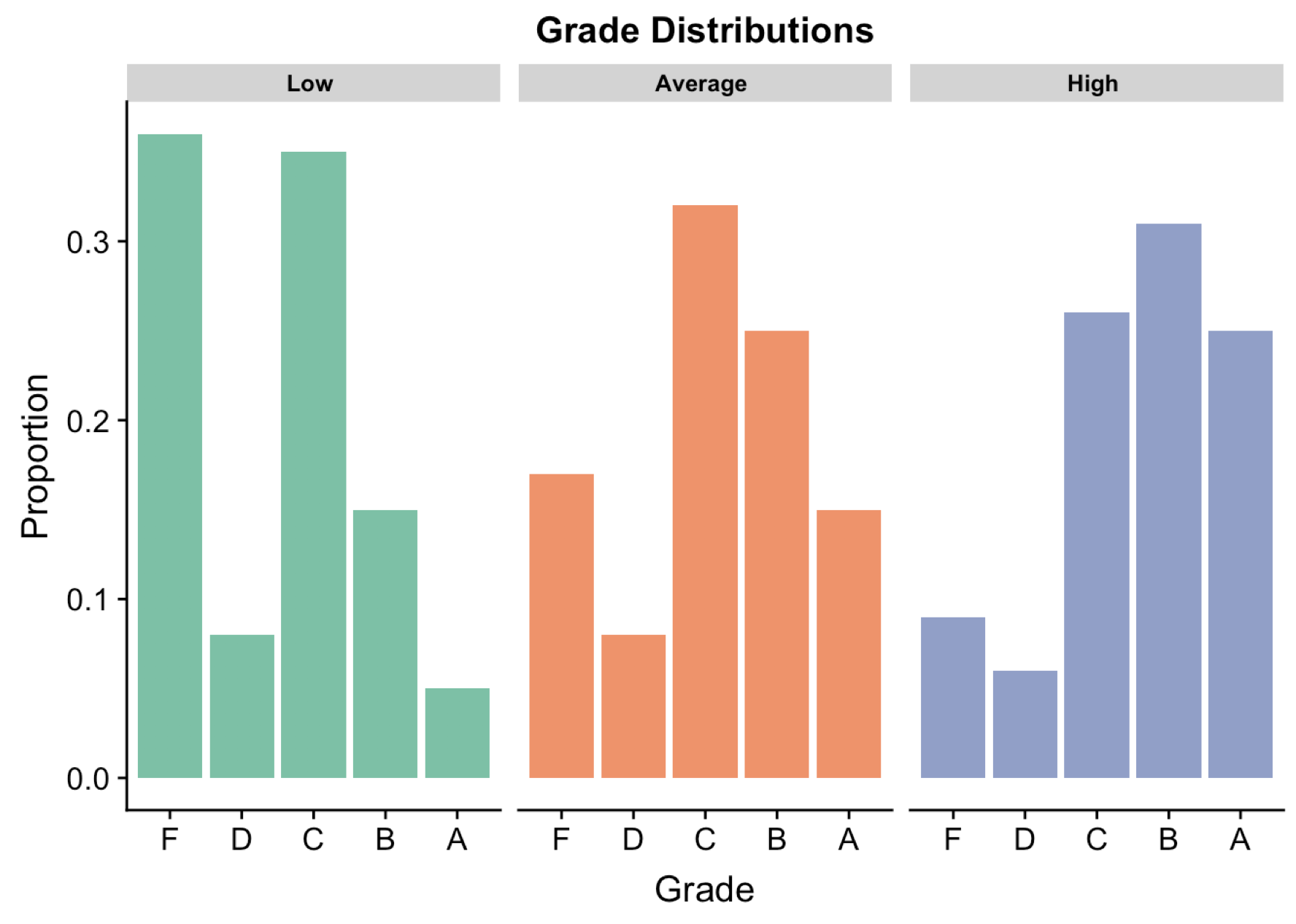}
\caption{Three grade distributions based on grades from 192 STEM courses.}
\label{fig:gradedistributions}
\end{figure}

 To determine the number of students that earned each grade in our simulated courses, we analyzed grade distributions from 192 STEM courses at California State University - Chico to build three different grade distributions: low, average, and high. We combined the drop, withdraw, and fail grades into a single F group. To build the low grade distribution, we averaged the grade distributions from 13 courses with less than 10\% As and greater than 30\% Fs. We built the average grade distribution by averaging all 192 grade distributions.  To build the high-grade distribution, we averaged the grade distributions from 6 courses with greater than 20\% As and greater than 20\% Bs. Fig.~\ref{fig:gradedistributions} shows the three grade distributions. We reasoned that these three distributions covered the range of grade distributions found in most STEM courses. 
 \par We simulated courses based on a course size of 1,000 students.  While this size is larger than typical courses, it allowed us to use fewer replications (twenty) of the course level simulations to quantify any bias introduced by MI or complete-case analysis. The actual size of each simulated course was 990 for the low grade distribution and 970 for the medium and high grade distributions. These sizes differed from each other and from 1,000 due to rounding in the course grade data we used to calculate the three grade distributions.

%Figure 6

\subsubsection{Simulated course data} \label{sec:simulated courses}
The 5 performance models and three grade distributions created a total of 15 different simulated courses. For each of these 15 courses, we simulated 20 datasets (replications) with approximately 1,000 students each. This process resulted in 300 different datasets. 
\par
\begin{figure}
\includegraphics[width=1\columnwidth]{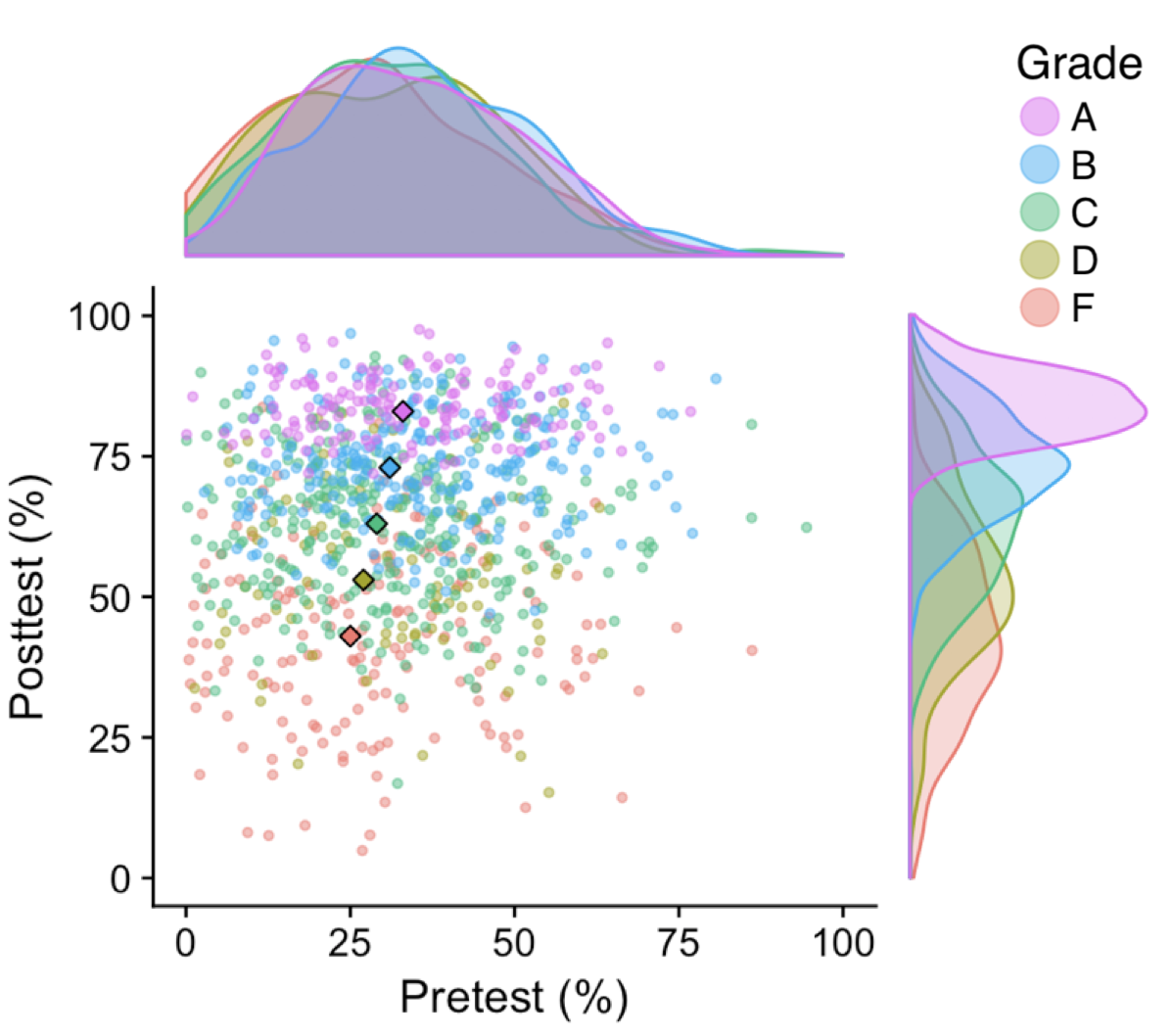}
\caption{Example data for an average grade distribution and high slope performance model. The diamonds are located at the means for each grade and illustrate the linear relationship between grade and mean test score. The density plots display the marginal distributions of the simulated pretest and posttest data for this simulated course.}
\label{fig:generated_data}
\end{figure}

 	Figure \ref{fig:generated_data} provides an example of data generated for one course using the high slope model with an intercept of 43 and a slope of 10 for the posttest scores and an average grade distribution. For the high slope model, each grade higher increased the average posttest concept inventory score by 10 percentage points. Students with F grades had a 43\% posttest score on the concept inventory on average and this raised to 53\% for Ds, 63\% for Cs, 73\% for Bs, and 83\% for As. The diamonds in Fig.~\ref{fig:generated_data} represent the mean test scores for the subgroups and illustrate the linear relationship between grade and both pretest and posttest means. The density plots for the pretests (top of Fig.~\ref{fig:generated_data}) and posttest (right of Fig.~\ref{fig:generated_data}) illustrate the variance of the generated scores about the means. The density plots for posttest scores covered a larger range of means and illustrate how the quadratic equation for standard deviation concentrated the scores into a narrower range as the mean score neared 100\%.
\par 	Table \ref{tab:models} provides the true average values for the complete data for pretest and posttest means and the absolute gain across the simulated courses. 
%Figure 7

%Table3*
\begin{table}[t]
\caption{Descriptive statistics for the 15 simulated courses average true pretest and posttest scores and gains.}\label{tab:models}
{\scriptsize
\begin{tabular}{ccccccc}\hline\hline\noalign{\smallskip}
Perfor-&Inter-&Slope&Grade & $\mu_{\mathrm{pre}}$ &$\mu_{\mathrm{post}}$ & Gain\\
mance&cept&		&Dist.&(\%)&(\%)& (\%)\\
\hline\noalign{\smallskip}
\multicolumn{7}{c}{\underline{Average Model}}\\ \noalign{\smallskip}
Average&43&6	&Low	&30.2&51.7&21.5\\
Average&43&6	&Average	&31.4&55.9&24.5\\
Average&43&6	&High	&32.1&58.5&26.4\\\noalign{\smallskip}
\multicolumn{7}{c}{\underline{Changing Intercept Models}}\\ \noalign{\smallskip}
Low Int. &25&6		&Low	&30.2&47.5&17.3\\
Low Int. &25&6		&Average	&31.4&49.4&18.0\\
Low Int. &25&6		&High	&32.1&50.8&18.7\\\noalign{\smallskip}
High Int. &58&6	&Low	&30.2&57.5&27.2\\
High Int. &58&6	&Average	&31.4&64.3&32.8\\
High Int. &58&6	&High	&32.1&68.9&36.8\\ \noalign{\smallskip}\noalign{\smallskip}
\multicolumn{7}{c}{\underline{Changing Slope Models}}\\ \noalign{\smallskip}
Low Slope &43&3		&Low	&30.2&35.3&5.1\\
Low Slope &43&3		&Average	&31.4&38.8&7.4\\
Low Slope &43&3		&High	&32.1&41.2&9.1\\\noalign{\smallskip}
High Slope &43&10	&Low	&30.2&66.6&36.3\\
High Slope &43&10	&Average	&31.4&70.7&39.2\\
High Slope &43&10	&High	&32.1&73.5&41.4\\
\noalign{\smallskip} \hline\hline
\end{tabular}
}
\end{table}

\subsection{Models for missing data} \label{sec:models for missing data}
We used the participation models for computer-based posttests from \citet{Nissen2018b} to create five levels of MAR data based on course grades in the simulated posttest data for each of the 15 simulated courses described in Table \ref{tab:models}. Table \ref{tab:partrats} and Fig.~\ref{fig:lonestar} show the five models for missing data with the value for `recommended practices' distinguishing between the five models. We used the model predictions provided by \citet{Nissen2018b} that used the average value for gender because we did not include gender as a variable in our simulated data.

\par To insert missing data into the posttest scores, first, we dissagregated the simulated complete data by course grade. Then, we used the participation models to determine the number of posttest scores that should be missing for that grade according to that model. Finally, we randomly deleted the appropriate number of posttest scores. As an example, for participation Model 2, Table \ref{tab:partrats}, (i.e., recommended practices = 2) we deleted 96\% of posttest scores for Fs, 83\% for Ds, 51\% for Cs, 18\% for a Bs, and 4\% for As. The randomization for deleting the posttest scores was done independently across all simulated datasets. Removing posttest scores represents a typical situation in which a student withdraws from the course or decreases their participation in the course at the end of the semester. Removing only posttest scores had a limited impact on the complete-case analysis because complete-case analysis removes both pretest and posttest scores when either is missing. These methods for generating missing data provided participation rates, the percentage of students who took both the pretest and posttest, that covered the range of 30\% to 80\% reported in the literature and presented in Table \ref{tab:lit review}. 

%Table4a
\begin{table}[t]
\caption{Participation rates for each final course grade based on models from \citet{Nissen2018b}. The model number represents the number of recommended practices to maximize student participation input into the final model.}\label{tab:partrats}
{\scriptsize
\begin{tabular}{lccccc }\hline\hline\noalign{\smallskip}
Grade & Model 0 & Model 1 & Model 2 & Model 3 & Model 4\\\hline\noalign{\smallskip}
A   & 0.30  & 0.75  & 0.96 & 0.99 & 1.00\\
B   & 0.13  & 0.45  & 0.82 & 0.96 & 0.99\\
C   & 0.05  & 0.18  & 0.49 & 0.81 & 0.95\\
D   & 0.02  & 0.05  & 0.17 & 0.41 & 0.71\\
F   & 0.01  & 0.02  & 0.04 & 0.10 & 0.24\\
\noalign{\smallskip} \hline\hline
\end{tabular}
}
\end{table}

%Figure 8
\begin{figure}
\includegraphics[width=1\columnwidth]{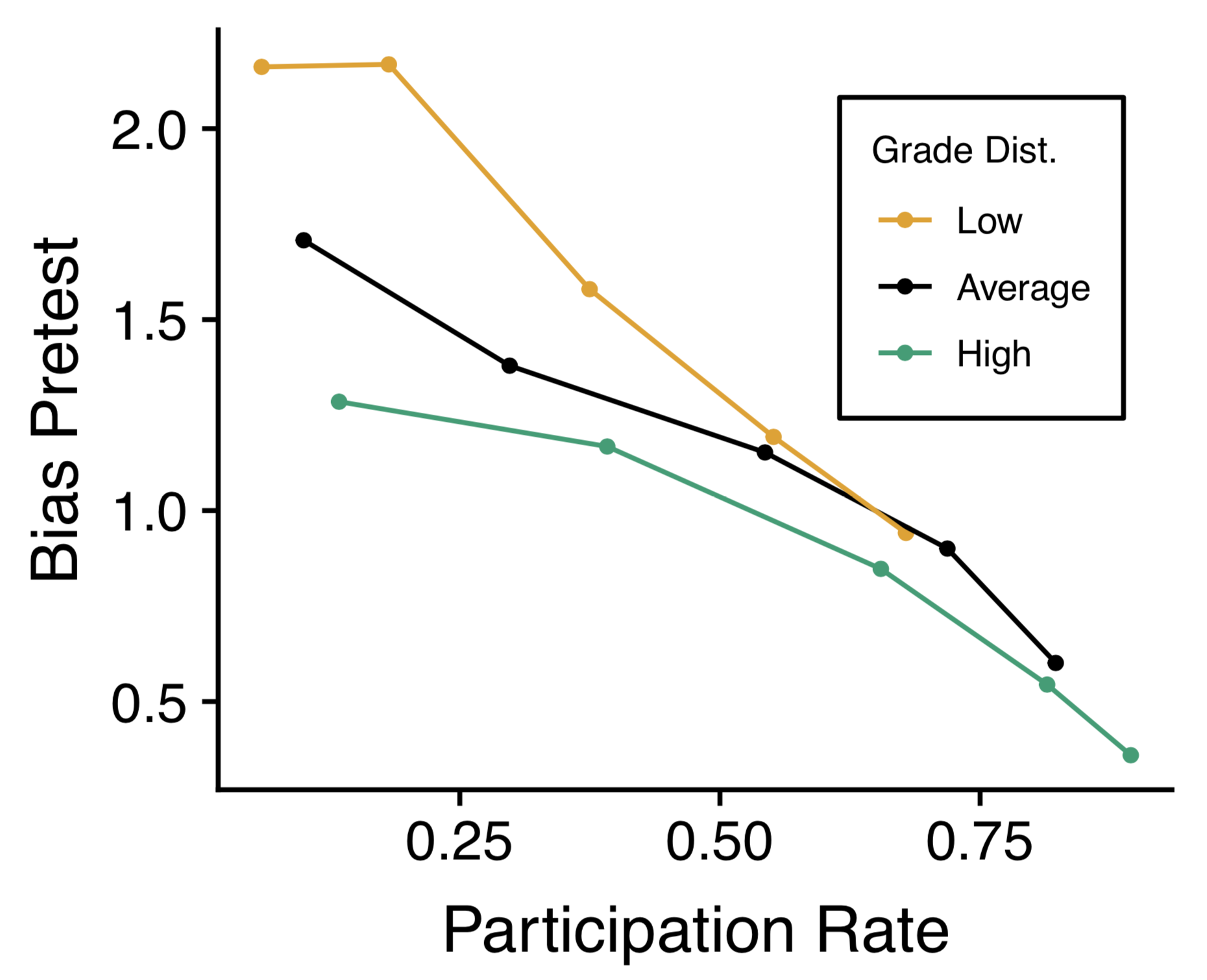}
\caption{Bias in the pretest model for the three grade distributions. Only the bias for the complete-case analysis are presented because no data was missing for the pretest and therefore the MI estimates could not be biased.}
\label{fig:pretest}
\end{figure}

%Figure 9
\begin{figure*}
\includegraphics[width=1\linewidth]{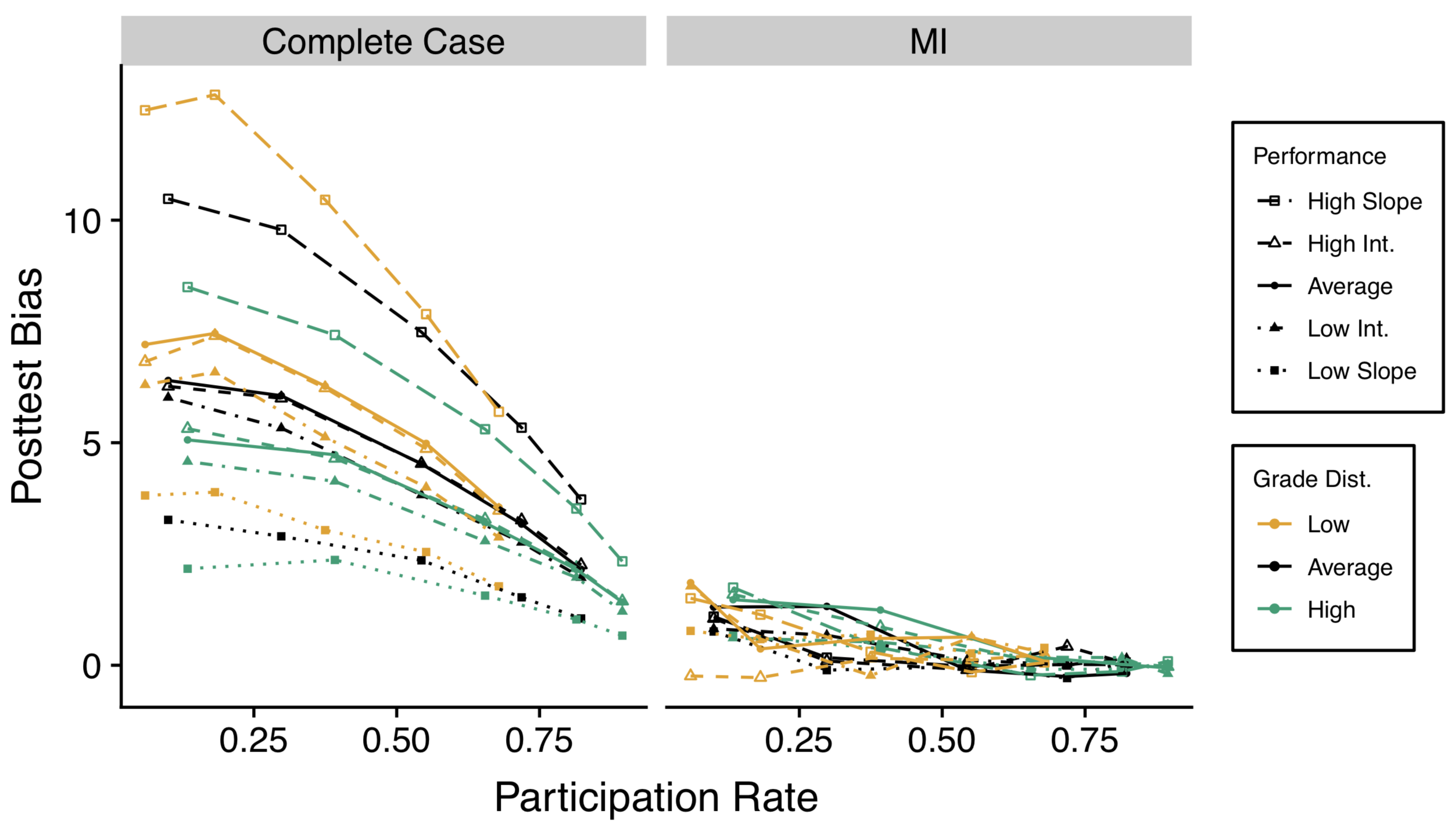}
\caption{Bias in the posttest data introduced by complete-case analysis or MI.}
\label{fig:posttest_bias}
\end{figure*}

\subsection{Measuring accuracy using bias}
\par 	To inform the extent to which complete-case analysis and MI provided biased estimates for posttest scores, we measured the accuracy of the results using bias. We calculated bias as the average difference between the true posttest mean ($\mu$) and the mean from either the complete-case or MI analysis ($\hat{\mu}$). This formula is shown in Eq.~(\ref{eq:6}) where $n$ represents the number of replications, which we set at 20 for each of the simulated courses. A bias greater than zero indicated that the estimates were larger than the true values.

\begin{equation}
\label{eq:6}
bias = \frac{1}{n}\sum_{i=1}^{n} \hat{\mu}_i-{\mu}_{i}.
\end{equation}

\section{Results}
We first present the bias on the pretest model across the three grade distributions. Second, we present the bias in the posttest scores for the 15 simulated courses. Last, we present a comparison of two simulated courses to illustrate the potential impact of the bias introduced by complete-case analysis and MI on research results.

\par 	We used the same model of the relationship between grade and test scores to simulate the pretest data for all five of the performance models because we expected the bias for the estimates of pretest scores to be smaller than that for the posttest scores. Figure \ref{fig:pretest} presents the pretest bias introduced by complete-case analysis. The participation models only inserted missing data in the posttests. The complete-case analysis created missing pretest data by discarding the pretest scores from students that do not participate in the posttest. MI discards no data and there were no missing pretest scores so it introduced no bias into the analysis for the pretest scores. Complete-case analysis introduced small amounts of bias ($<2.2$ percentage points) into the course means for all  simulated datasets. The bias introduced by complete-case analysis for the pretest tended to increase as the participation rate decreased and tended to be higher for lower grade distributions.

\par 	The  posttest bias, shown in Fig.~\ref{fig:posttest_bias}, resulting after conducting complete-case analysis and MI tended to be positive and to overestimate the true values. Conducting complete-case analysis resulted in more bias than conducting MI. Conducting complete-case analysis always produced positive biases with a minimum value of 0.7 percentage points and a maximum value of 12.8 percentage points. The bias of 12.8 percentage points meant that complete-case analysis estimated the posttest mean to be 70.2\% on average for the high slope low grade distributions simulated course while the true average value was 57.4\%. In contrast to complete-case analysis, conducting MI produced negative biases for 19 of the 75 measurements with a minimum value of -0.3 percentage points and a maximum value of 1.9 percentage points. These results indicate that both methods tend to overestimate the true posttest scores, but that the overestimation was much larger for complete-case analysis. This overall trend of larger bias resulting from complete-case analysis than from MI was true for all 75 combinations of performance, grade distribution, and participation rates. Even at the lowest level of participation, the MI analysis tended to produce less bias than the highest level of participation for the complete-case analysis, as is illustrated by the boundary between the two graphs in  Fig.~\ref{fig:posttest_bias}. 
\par 	The bias introduced by conducting both MI and complete-case analysis tended to decrease as the participation rate increased. This trend occurred for complete-case analysis of all 15 of the simulated courses but was less consistent for MI analysis of the simulated courses. These results illustrate the value of maximizing participation rates for achieving accurate estimates of concept inventory means.
\par 	Differences in bias across the five performance models for complete-case analysis indicated that varying slope had a stronger impact on bias than varying intercept. As shown in Fig.~\ref{fig:posttest_bias}, the largest bias occurred for the high slope simulated courses (long-dashed line with empty squares) and the lowest bias occurred for the low slope simulated courses (dotted line with filled squares). The maximum bias for the high-slope simulated courses was 12.8 percentage points whereas the maximum bias for the high-intercept simulated courses (dashed lines with empty triangles) was 7.4 percentage points. This difference in bias was not caused by a difference in posttest scores as the bias was larger in the high-slope simulated courses but the mean posttest score was lower (57.4\% for the 12.8 percentage point bias versus 66.6\% for the 7.4 percentage point bias). Similarly, comparing the low slope and low intercept high grade distribution simulated courses shows that the bias for the low slope course was lower (0.7 versus 1.2 percentage points maximum bias for each). Whereas, the posttest mean was higher for the low-slope simulated courses (50.7\% for 0.7 percentage point bias versus 41.9\% for 1.2 percentage point bias). These relations indicated that the absolute value of the posttest mean was not the primary factor in the amount of bias introduced by complete-case analysis. Rather, the relationships within the datasets and the total amount of missing data best explained the bias. 
\par 	Unlike complete-case analysis, the bias for MI did not reveal consistent differences between the performance models or grade distributions and bias. The much lower overall bias for MI may obscure differences in bias across the simulated courses. However, Fig.~\ref{fig:posttest_bias} shows that the clear differences in bias for complete-case analysis across the simulated courses did not exist for MI.

%Figure 10
\begin{figure}
\includegraphics[width=1\columnwidth]{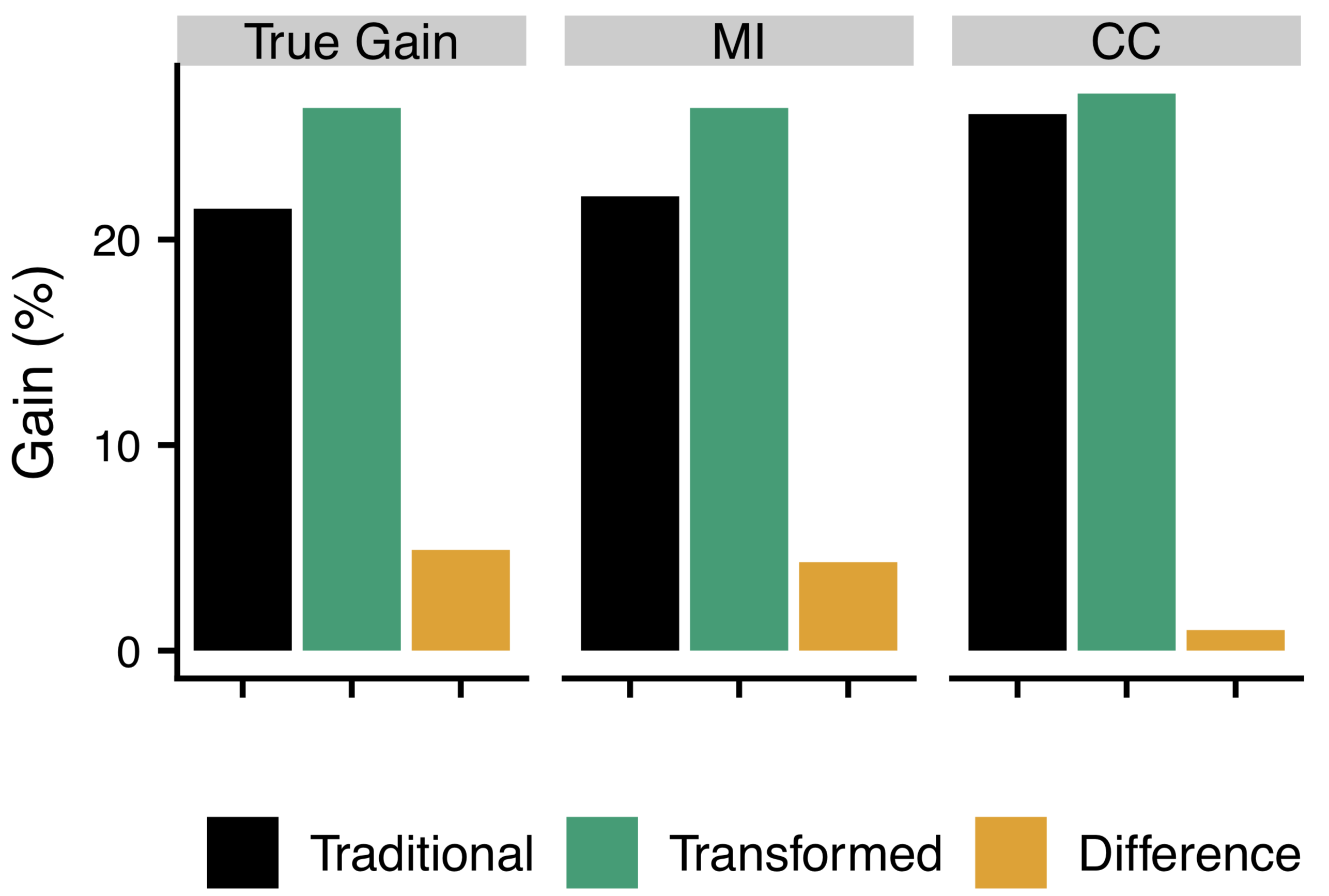}
\caption{Bar graph illustrating the effect of bias from complete-case analysis or MI on a comparison of two courses. Performance in both courses was average. The traditional course had a low grade distribution and low participation rates. The transformed course had a high grade distribution and a high participation rate. We did not include error bars to focus on the effects of bias and because they are very small due to the large sample sizes for the simulated data.}
\label{fig:example}
\end{figure}

\par 	To compare how the bias introduced by complete-case analysis and MI could skew comparisons, in Fig.~\ref{fig:example} we compared two simulated courses with similar performance within each grade but different grade distributions and participation rates. Using the average performance model for both courses simplified comparing the results because the performance for students who earned the same grade were the same across the two courses.  We varied the participation and grade distributions between the two courses to align with comparisons between traditional and transformed courses that occur in the PER literature (e.g., \citet{Brewe2010}).  The two comparison courses are listed below.
\begin{enumerate}
\item Traditional Course
 \begin{enumerate}
 \item Average performance within each grade
 \item Low grade distribution
 \item Low participation (37\%)
 \end{enumerate}
\item Transformed Course
 \begin{enumerate}
 \item Average performance within each grade
 \item High grade distribution
 \item High participation (81\%)
 \end{enumerate}
\end{enumerate}

\par 	The true values indicated that students in the transformed course learned more conceptual knowledge on average than the students in the traditional course. This difference follows from the higher grade distribution in the transformed course and the same performance model in both courses. The larger gains in the transformed course remained when we analyzed the data with MI. However, complete-case analysis nearly eliminated the difference in gains on the concept inventory. This decrease in the difference between the courses occurred because little data was collected in the traditional courses from students with low grades and thus the analysis positively biased the gain. In contrast to the true results and the results after analysis with MI, the results from the complete-case analysis do not support the claim that students learned substantially more in the transformed course than in the traditional course.

\section{Discussion}

Complete-case analysis can introduce large amounts of bias into the estimates for concept inventory scores when researchers apply it to data that is not MCAR. The bias introduced by complete-case analysis in the simulated data ranged from 0.7\% to 12.8\% for the posttest means and fell below 2\% for the pretest means. The 28 articles we reviewed, which included 158 courses, reported gains from 5\% to 56\% with an average of 23\%. Twenty three of these studies used complete-case analysis, none reported using a principled method for handling missing data (e.g., MI), and none indicated that the missing data in the study was MCAR. Subsequently, our results indicate that part of the gains reported in those studies likely resulted from the improper use of complete-case analysis. In some of those studies, complete-case analysis may have exaggerated the gains by increasing them from anywhere between one third to doubling them. The introduced bias may have also skewed any comparisons made in those studies, particularly comparisons across courses with different participation rates.
\par 	We cannot say exactly how much of these reported gains resulted from bias introduced by complete-case analysis. Our results indicate that the amount of bias complete-case analysis introduces depends on both the participation rate and the relationships within the data. To determine the bias in prior studies that used complete-case analysis without meeting the assumptions for its reliable use, researchers will need to analyze the data directly. However, physics education researchers seldom publish the data or analytical code used in their studies. The PER community can improve transparency and accountability by supporting researchers in publishing or publicly sharing the datasets from their research. Going forward, sharing data would allow the research community to double check the impact that the methods for handling missing data have on the conclusions that researchers draw from their data. 
\par 	The bias introduced by complete-case analysis could obscure differences across courses and undermine both research and evaluation work. For example, we compared a simulated traditional course with a simulated transformed course. The simulated transformed course had lower DWF rates, higher grades, and greater conceptual learning. Bias introduced by using complete-case analysis obscured the differences in conceptual learning between the two simulated courses. In a comparison of real courses, a critic of the transformed course with lower DWF rates could use the similar results from the complete-case analysis of the concept inventory scores to claim the transformed course had lower grading standards. Otherwise, the transformed course would have outperformed the other course on the concept inventory. Using MI to account for the missingness in the data introduced less bias into the results and preserved the true result that, overall, students learned more in the transformed course. Researchers and educators need accurate results to inform the design and implementation of research-based teaching materials. If researchers continue to use complete-case analysis without accounting for the impact of missing data, they risk wasting time and resources either discarding useful interventions or pursuing false leads.

\section{Conclusion}

\par 	Researchers, reviewers, and editors can take several steps to improve the handling of missing data in quantitative studies. During the data collection process, researchers should take reasonable actions to minimize the amount of missing data. However, education researchers often cannot avoid some missing data in their studies. Researchers should use MI or another principled method for handling missing data. Researchers using complete-case analysis should present evidence that their data is MCAR. However, principled methods for handling missing data, such as MI, are not a panacea. Rather, principled methods are only one component of the diligence necessary to address missing data. Before analyzing the data and deciding on an appropriate method for handling the missing data, researchers should examine the amount of missing data; patterns in the missing and complete data; and the mechanisms behind those patterns. When implementing MI to address missing data, researchers should check that their data meets the assumptions of the MI algorithm. Many MI software packages include tools to check these assumptions. Studies should state the participation rates in their data collection, describe the methods they used to address missing data, discuss patterns in the missing data, and discuss how the missing data may influence analytical results. These steps will improve the quality, reliability, and replicability of quantitative studies on student outcomes in physics.

\section {Acknowledgements}
\par This work is funded in part by NSF-IUSE Grant No. DUE-1525338 and is Contribution No. LAA-059 of the Learning Assistant Alliance. We are grateful to the Learning Assistant Program at the University of Colorado Boulder for establishing the foundation for LASSO and LASSO studies.

\bibliography{MI.bib}

\end{document}